\documentclass[
    ,final            
  ]
  {aipproc}

\layoutstyle{6x9}

\begin{document}

\title{Stress-free BCS pairing in color superconductors is impossible}


\author{Krishna Rajagopal}{
  address={Center for Theoretical Physics, Massachusetts Institute of Technology, Cambridge, MA 02139}
}

\author{Andreas Schmitt}{
  address={Department of Physics, Washington University St Louis, MO 63130}
}

\begin{abstract}
Cold, asymptotically dense three-flavor quark matter is in the color-flavor locked (CFL) 
phase, in which all quarks pair in a particularly symmetric fashion. 
At smaller densities, taking into account a nonzero strange quark mass and electric and
color neutrality, the CFL phase requires pairing of quarks with mismatched Fermi momenta.
We present a classification of all other possible, less symmetric, pairing patterns and prove
that none of them can avoid this mismatch. This result suggests unconventional, e.g., 
spatially inhomogeneous, superconducting phases for moderate densities.
\end{abstract}

\maketitle


At asymptotically large densities and sufficiently cold temperatures, three-flavor quark matter 
is in the color-flavor locked (CFL) phase \cite{Alford:1998mk}. This phase is a color
superconductor because the color gauge group $SU(3)_c$ is spontaneously broken, due to the formation of quark 
Cooper pairs. The underlying mechanism is an attractive QCD interaction betweeen the quarks that are
antisymmetric in color. Consequently, the order parameter is a color antitriplet and, due to 
the overall antisymmetry of the two-fermion wave function, also an antitriplet in flavor space
(assuming pairing in the spin-0 channel). Hence we can write the order parameter as
\begin{equation}
{\cal M} = \Delta_{ij} J_i\otimes I_j \, , 
\end{equation}
with the antisymmetric color and flavor matrices $(J_i)_{jk}=(I_i)_{jk}=-i\epsilon_{ijk}$. The 
complex $3\times 3$ matrix $\Delta$ determines the pairing pattern. In the CFL phase, 
$\Delta_{ij} \propto \delta_{ij}$, and all nine quasiquarks  
acquire a gap in their excitation spectrum. The reason is that ${\cal M}{\cal M}^\dagger$ 
has nine nonzero eigenvalues. 

At large densities the quark chemical potential $\mu$ is much  
larger than all three quark masses and we may approximate $m_u\simeq m_d\simeq m_s\simeq 0$.
In this case, the Lagrangian of the system is invariant under color and flavor transformations
$SU(3)_c\times SU(3)_f$. Therefore, any order parameter $\Delta$ is equivalent to rotated 
order parameters $U^T\Delta V$, with $U\in SU(3)_c$, $V\in SU(3)_f$. It is thus sufficient 
to consider diagonal matrices $\Delta$, and it is easy to see that the preferred phase
(= lowest in free energy) is the CFL phase.

Things get more complicated for moderate densities. For instance, in the interior of compact stars
we expect the quark chemical potential to be of the order of 500 MeV at most. In this case, 
we may not neglect the strange mass anymore. For our purpose this has two important consequences. First, 
color and electric neutrality become important nontrivial constraints for the system
(for $m_s=0$ these constraints are trivially fulfilled in the CFL phase). In the CFL phase, 
the neutrality conditions in the case of a small but nonzero strange mass lead to pairing of 
quarks with mismatched Fermi surfaces. The difference in chemical potentials is of the order
$m_s^2/\mu$. This imposes a stress on the pairing. A cost in free energy of the order of $m_s^4$
has to be paid in order to fill both 
participating quark states up to a common Fermi momentum $\nu$. Only then, conventional BCS pairing
with zero-momentum Cooper pairs can be achieved, gaining condensation energy of the order of 
$\Delta^2\mu^2$, $\Delta$ being the energy gap in the quasiparticle spectrum.
At some particular value of the parameter $m_s^2/\mu$, the cost exceeds the gain and conventional
pairing is no longer favored. For the CFL phase, this value is $m_s^2/\mu = 2\Delta$.

The second important consequence of a nonzero strange mass is the explicit breaking of the flavor 
symmetry down to $SU(2)_f$. There exist not necessarily transformations $U\in SU(3)_c$ and $V\in SU(2)_f$, 
which diagonalize a given complex matrix $\Delta$. Hence, there are non-diagonal order parameters 
which describe physically distinct phases. Such order parameters are known from systems
with similar mathematical structure, cf.\ the {\it A}-phase in a 
spin-1 color superconductor \cite{Schmitt:2004et} or in $^3$He. 
An interesting 
question arises regarding the fate of these phases at moderate densities. While it is clear that 
the CFL phase is favored for large $\mu$, for smaller $\mu$ there might be a phase which is more comfortable 
with the neutrality constraints. More precisely, we may ask whether there is some less symmetric,
but still conventional pairing pattern in which, once electric and color neutrality are 
imposed, pairing only occurs among those quarks whose Fermi momenta would be equal in the 
absence of pairing.

In order to answer this question, we have to set up formal conditions for $(i)$ electric and color
neutrality and $(ii)$ stress-free pairing, and have to find a way to exhaust all possible pairing patterns
systematically. Conditions $(i)$ and $(ii)$ are derived upon introducing an electric chemical 
potential $\mu_e$ and color chemical potentials $\mu_3$ and $\mu_8$. In general, a color-superconducting
phase may require additional or other nonzero color chemical potentials. For our purpose,
$\mu_3$ and $\mu_8$ are sufficient since any pattern can be transformed into one where only 
$\mu_3$ and $\mu_8$ are nonzero \cite{Rajagopal:2005dg}. Then, 
neglecting higher order terms in $m_s,\Delta\ll\mu$, conditions 
$(i)$ and $(ii)$ are \cite{Rajagopal:2005dg}
\begin{equation}\label{conditions}
0 = \sum_{i=1}^9\nu_i q_{ik} \, , \qquad \nu_i = \mu_i^{\rm eff} \, .
\end{equation}
Here, $\nu_i$ are the common Fermi momenta, assigned to each of the nine 
quarks according to their pairing partners. Moreover, $k=e,3,8$, and 
$\mu_i^{\rm eff}=\mu_i$ for up and down quarks and $\mu_i^{\rm eff}=\mu_i-m_s^2/(2\mu)$
for strange quarks, where $\mu_i = \mu + q_{ie}\mu_e + q_{i3}\mu_3 + q_{i8}\mu_8$. The 
charges of the quarks are denoted by $q_{ik}$. 

We observe that the only information we need to know about a particular pairing pattern are 
the common Fermi momenta $\nu_i$ as a function of $\mu,m_s,\mu_e, \mu_3, \mu_8$.  
A common Fermi momentum is the arithmetic mean of the effective chemical potentials of the 
quarks that pair with each other \cite{Rajagopal:2005dg}. 
Consequently, all we need to know is which quark pairs with
which other quarks. This information is given by setting the 9 entries $\Delta_{ij}$ of the
order parameter to zero or nonzero (say 1) in all possible combinations. Hence, respecting the 
symmetries given by
the structure of ${\cal M}$, we have to investigate $2^9=512$ patterns. One of these, $\Delta_{ij}=0$
for all $i,j$, corresponds to unpaired quark matter. 
Note that many of the 512 patterns are related by color rotations,
meaning that not all of them are physically distinct.   

In order to automatize the treatment of all patterns, we have translated each pattern into
a graph. The vertices of this graph are the nine quarks while there are at most $9\cdot 2=18$ 
edges, depending on the matrix $\Delta$. Every nonzero entry $\Delta_{ij}$ yields two edges, 
meaning that the corresponding quarks form Cooper pairs with each other. Then, a pattern is, 
for our purposes, given by the connected components of the graph. Any pattern uniquely defines its set
of components while any given set of components may corresponds to several patterns; it turns out
that the 512 patterns yield 149 distinct sets of components. Once the patterns are classified
in this way, we may use a computer program to test each set of components on the conditions
(\ref{conditions}): we first determine the common Fermi momenta for each set and then search for a 
simultaneous solution $(\mu_e, \mu_3, \mu_8)$ of Eqs.\ (\ref{conditions}). We find that none 
of the patterns allows for such a solution (except for unpaired quark matter). 
Consequently, {\em neutral stress-free BCS pairing is impossible}. 

We emphasize that this result does not exclude any less symmetric phase in the 
phase diagram. We rather conclude that any of these phases must break down at a point
in density that is parametrically the same as for the CFL phase and given by a quark chemical potential 
$\mu$ of the order of $m_s^2/\Delta$. If this value of $\mu$ is larger than the one at which the 
phase transition from quark matter to nuclear matter occurs, one can expect some form
of unconventional pairing which succeeds CFL down in density. Assuming the  
CFL pairing pattern and a spatially homogeneous system, the CFL phase 
is succeeded by a phase in which some quasiparticle excitations become gapless despite a nonzero
order parameter. This ``gapless CFL'' phase, however, is unstable with respect to 
the formation of counter-propagating currents. In the simplest situation, there are two opposite 
currents, provided by the Cooper pair condensate (or possibly a kaon condensate) 
and ungapped fermionic modes. In a more complicated version, crystalline structures arise
\cite{Rajagopal:2006ig}. Other possibilities have been proposed, for instance single-flavor pairing
\cite{Schmitt:2004et}. Most of the proposed phases have ungapped modes. Since transport
properties (e.g., neutrino emissivity) are very sensitive to these ungapped modes \cite{Schmitt:2005wg}, 
related astrophysical observables (e.g., cooling curves) might help in the search for the ground state.

\begin{theacknowledgments}
The authors acknowledge support by the U.S. Department of Energy
under contracts DE-FG02-91ER50628, DE-FG01-04ER0225 (OJI), and 
DF-FC02-94ER40818.
\end{theacknowledgments}


\end{document}